\documentclass{ametsocV6}
\usepackage[number-unit-product=~]{siunitx}

\newcommand{\Ek}{\text{Ek}}
\newcommand{\pder}[2]{\frac{\partial#1}{\partial#2}}
\renewcommand{\vec}[1]{\mathbf{#1}}

\title{Rapid Spin Up and Spin Down of Flow Along Slopes}
\authors{Henry G. Peterson\correspondingauthor{hgpeterson@caltech.edu} and J\"orn Callies}
\affiliation{California Institute of Technology, Pasadena, California}

\abstract{The near-bottom mixing that allows abyssal waters to upwell tilts isopycnals and spins up flow over the flanks of mid-ocean ridges.
Meso- and large-scale currents along sloping topography are subjected to a delicate balance of Ekman arrest and spin down.
These two seemingly disparate oceanographic phenomena share a common theory, which is based on a one-dimensional model of rotating, stratified flow over a sloping, insulated boundary.
This commonly used model, however, lacks rapid adjustment of interior flows, limiting its ability to capture the full physics of spin up and spin down of along-slope flow.
Motivated by two-dimensional dynamics, the present work extends the one-dimensional model by constraining the vertically integrated cross-slope transport and allowing for a barotropic cross-slope pressure gradient. 
This produces a closed secondary circulation by forcing Ekman transport in the bottom boundary layer to return in the interior.
The extended model can thus capture Ekman spin up and spin down physics: the interior return flow is turned by the Coriolis acceleration, leading to rapid rather than slow diffusive adjustment of the along-slope flow.
This transport-constrained one-dimensional model accurately describes two-dimensional mixing-generated spin up over an idealized ridge and provides a unified framework for understanding the relative importance of Ekman arrest and spin down of flow along a slope.}

\begin{document}

\maketitle

\section{Introduction}

The ocean is a rotating, stratified shell of fluid with a geometrically complicated bottom boundary. 
The sloping seafloor affects a number of aspects of the ocean's circulation. 
It allows near-bottom diapycnal mixing to bend isopycnals and thus spin up a circulation in the abyss \citep[e.g.,][]{phillips_flows_1970,wunsch_oceanic_1970,garrett_role_1990,callies_dynamics_2018}, and it allows for bottom Ekman layers to be arrested by buoyancy forces and thus for currents to slide along slopes without being spun down \citep{rhines_boundary_1989,maccready_buoyant_1991,maccready_slippery_1993}. 
These spin up and spin down processes have long been studied using the equations of motion in a coordinate frame that is rotated to align with the sloping bottom and simplified by considering variations in the slope-normal direction only \citep[see][for a review]{garrett_boundary_1993}.
We here argue that new insight can be gained by enforcing a transport constraint in these one-dimensional dynamics and allowing for a time dependent cross-slope barotropic (vertically constant) pressure gradient. 
These modifications enable boundary mixing to spin up an interior flow much more rapidly than through ``slow diffusion'' \citep{maccready_buoyant_1991}, and they allow for Ekman spin down in addition to Ekman arrest, capturing the competition between the two processes.

Enhanced turbulent mixing near the seafloor is thought to be a crucial element of the overturning circulation of the abyssal ocean, and 1D dynamics have been a powerful tool for understanding the dynamical response to such mixing over a sloping bottom.
Antarctic Bottom Water fills the abyss of the Atlantic and Pacific basins \citep[e.g.,][]{lumpkin_global_2007,talley_closure_2013}.
For these dense waters to return to the surface, they must cross isopycnals and thus require diapycnal mixing \citep[e.g.,][]{munk_abyssal_1966,munk_abyssal_1998,ferrari_what_2014}.
Observations have revealed that this diapycnal mixing is strongly enhanced over rough topography \citep[e.g.,][]{polzin_spatial_1997,ledwell_evidence_2000,waterhouse_global_2014}, where tidal and geostrophic currents produce a field of vigorous internal waves that break and produce small-scale turbulence \citep[e.g.,][]{garrett_internal_2007,nikurashin_global_2011}.
Our understanding of how the ocean responds to this mixing, both locally and globally, has been shaped by 1D theory for a stratified, rotating fluid overlying a sloping, insulated seafloor \citep[e.g.,][]{phillips_flows_1970,wunsch_oceanic_1970,thorpe_current_1987,garrett_boundary_1993}.
This theory (and the thinking it inspires), suggests that bottom-intensified mixing spins up diabatic upslope flow in a thin bottom boundary layer and diabatic downslope flow in a stratified mixing layer above \citep{garrett_role_1990,ferrari_turning_2016,de_lavergne_consumption_2016,mcdougall_abyssal_2017,callies_restratification_2018}. 
Variations in these locally produced flows give rise to exchange with the interior and produce a basin-scale circulation in the abyss \citep[e.g.,][]{phillips_experiment_1986,mcdougall_dianeutral_1989,garrett_marginal_1991,dell_diffusive_2015,callies_dynamics_2018,drake_abyssal_2020}.

\begin{figure}
    \centering
    \includegraphics[width=19pc]{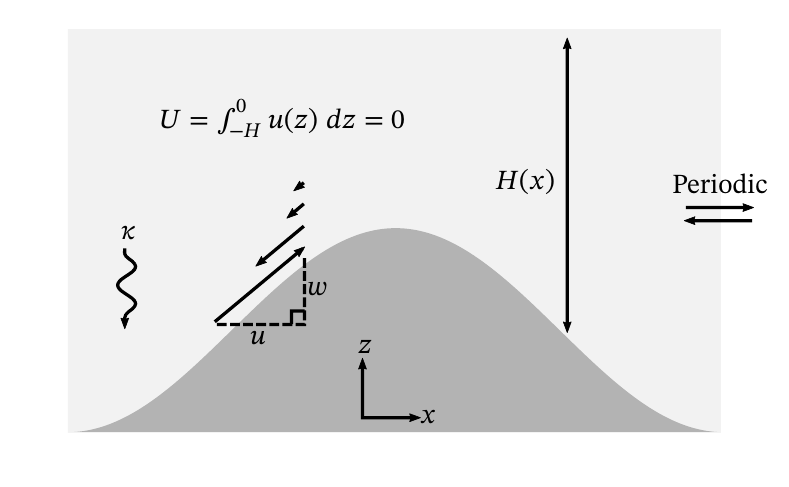}
    \caption{Sketch of idealized mid-ocean ridge geometry.
    By continuity and symmetry, the vertically integrated cross-ridge transport $U$ must vanish.}
    \label{fig:sketchRidge}
\end{figure}

It has recently become clear, however, that the canonical 1D theory falls short in capturing two- and three-dimensional abyssal spin up, even in highly idealized contexts.
The cross-slope mean flow generated by the 1D system is too weak to keep abyssal mixing layers stratified and instead produces a configuration that is baroclinically unstable \citep{wenegrat_submesoscale_2018,callies_restratification_2018}.
Even if the role of baroclinic eddies is set aside, as will be done in the remainder of this work, spin up in two dimensions is qualitatively different from that predicted by 1D theory.
\citet{ruan_mixing-driven_2020}, considering bottom-intensified mixing over an idealized mid-ocean ridge (cf.,~Fig.~\ref{fig:sketchRidge}), found that an interior flow along the ridge spins up rapidly, in direct contrast to the slow diffusion predicted by the canonical 1D equations \citep{maccready_buoyant_1991}.
We show below that this rapid adjustment can be captured in 1D dynamics if a constraint is imposed on the vertically integrated cross-slope transport.

The 1D dynamics have also been a cornerstone in our understanding of the spin down---or lack thereof---of meso- and large-scale geostrophic currents flowing along topographic slopes.
Over a flat bottom boundary, a current induces Ekman transport in the bottom boundary layer.
If the strength of the interior flow varies in the horizontal, so will the Ekman transport, leading to Ekman pumping and suction.
By continuity, this generates a secondary circulation so that the boundary layer transport is returned in the interior.
The Coriolis acceleration then turns the flow, spinning down the original current on a time scale of $\tau_S = f^{-1}\Ek^{-1/2}$ where $f$ is the inertial frequency, $\Ek = \nu/f H^2$ is the Ekman number, $\nu$ is a turbulent viscosity scale, and $H$ is a height scale  \citep[e.g.,][]{pedlosky_geophysical_1979}.
The sloping boundary adds new physics to the problem: as fluid is moved up- or down-slope due to Ekman transport, it experiences a buoyancy force that opposes its motion \citep{rhines_boundary_1989,maccready_buoyant_1991}.
If a balance between the Coriolis and buoyancy forces is reached, the Ekman transport is ``arrested.''
This shuts down the secondary circulation and halts further spin down, so that from then on the far-field current experiences an approximately free-slip bottom boundary condition.
The timescale at which Ekman arrest occurs is roughly $\tau_A = (Sf)^{-1}$ where $S = N^2\tan^2\theta/f^2$ is the slope Burger number for a fluid with buoyancy frequency $N$ over a slope at an angle $\theta$ above the horizontal \citep{maccready_buoyant_1991}.
The sloping topography thus enables the interior flow to persist if Ekman arrest is much faster than spin down, that is if $\tau_A/\tau_S = \Ek^{1/2}/S \ll 1$ \citep{garrett_boundary_1993}.

The canonical 1D model captures only the physics of Ekman arrest, not those of spin down.
In that model, the cross-slope Ekman transport produced by the initial along-slope flow need not be returned in the interior, so the secondary circulation that can spin down the along-slope flow is lacking.
While the physics of these two processes have now been known for decades, fully understanding their competition and interplay has been hampered by this disconnect.
\citet{chapman_deceleration_2002} captured both processes in a simplified bulk model, but the connection to the more complete 1D dynamics remained opaque.
We show below that a 1D model derived directly from the full equations of motion can capture the physics of spin down and arrest if a transport constraint is imposed and a cross-slope pressure gradient is included, the same two modifications to the canonical 1D dynamics that allow for a rapid adjustment of the interior flow in spin up.

\begin{figure*}
    \centering
    \includegraphics[width=33pc]{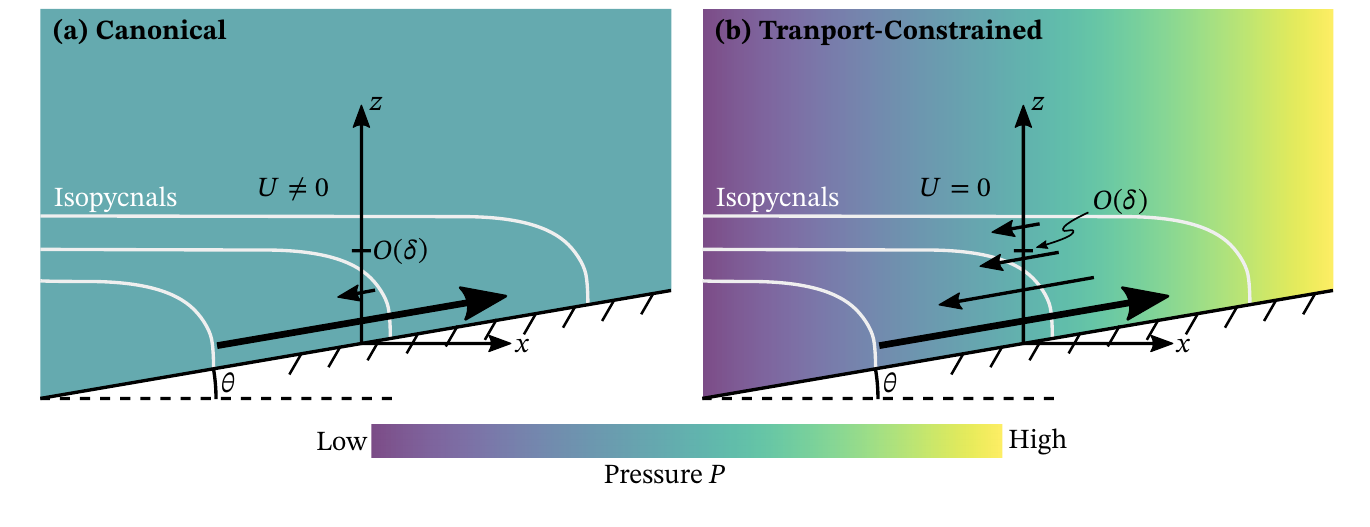}
    \caption{Difference between the canonical and transport-constrained 1D models. 
    Sketched are typical isopycnals and cross-slope flow $u$ as a function of $z$ for spin-up with constant mixing coefficients.
    Colors represent the barotropic pressure gradient $\partial_x P$.
    }
    \label{fig:sketchSlope}
\end{figure*}

The key innovation of this work, introduced more fully in the next section, is thus a transport-constrained 1D model capable of representing rapid spin up and spin down.
With the geometry sketched in Fig.~\ref{fig:sketchSlope} and using standard notation, the modified 1D dynamics are
\begin{align}
    \pder{u}{t} - f \varv &= -\pder{P}{x} + b \tan \theta + \pder{}{z} \left( \nu \pder{u}{z} \right) \label{eq:tc-x-intro},\\
    \pder{\varv}{t} + f u &= \pder{}{z} \left( \nu \pder{\varv}{z} \right) \label{eq:tc-y-intro},\\
    \pder{b}{t} + u N^2 \tan \theta &= \pder{}{z} \left[ \kappa \left( N^2 + \pder{b}{z} \right) \right] \label{eq:tc-b-intro}, \\
    \int_0^H u \; \text{d} z &= U \label{eq:tc-U-intro}.
\end{align}
Crucially, $U$ is an imposed cross-slope transport that we will typically set to zero, and $P$ is a barotropic pressure perturbation from the background state of rest.
The transport constraint enforces that any boundary layer transport must be returned outside the boundary layer, creating a secondary circulation and allowing for rapid adjustment of the along-slope flow.
It is possible to impose this transport constraint because we allow for an implicitly determined time-varying barotropic cross-slope pressure gradient $\partial_x P$.

In canonical 1D dynamics, this cross-slope pressure gradient is absent from~\eqref{eq:tc-x-intro} or fixed in time to balance an initial along-slope flow. 
In that case, an anomalous geostrophic flow~$\varv$ must satisfy the balance $-f \varv = b \tan \theta$ (which by hydrostatic balance equals $\partial_z p \tan \theta$, the projection of the vertical perturbation pressure gradient onto the slope). 
A change in the geostrophically balanced along-slope flow thus requires a typically slow modification of the buoyancy anomaly~$b$. 
The inclusion of a time-varying $\partial_x P$ in the modified equation~\eqref{eq:tc-x-intro} instead allows the balance $-f \varv = -\partial_x P$ and thus the rapid spin up or spin down of a (barotropic) geostrophic flow.

We derive this model in Section~\ref{s:theory}, where we motivate it by comparing the 1D and 2D equations in the planetary geostrophic (PG) limit.
We then demonstrate the utility of the modified model by considering mixing-generated spin up over an idealized mid-ocean ridge in Section~\ref{s:spinup} and the spin down of an along-slope current in Section~\ref{s:spindown}.
We offer a discussion in Section~\ref{s:discussion} and conclude in Section~\ref{s:conclusions}.

\section{Rapid Adjustment and Constrained Transport}\label{s:theory}

In this section, we motivate the transport-constrained 1D model summarized above.
We begin with a review of the canonical 1D theory, emphasizing the fact that it does not enforce any constraints on vertically integrated cross-slope transport.
We then find that, when considering the 1D and 2D systems in the PG limit, the inversion statements take the same form and include an explicit transport term.
In the 2D system, this term is constrained by the geometry of the domain.
With this in mind, we modify the 1D model to allow for constrained transport by including a time-varying barotropic pressure gradient term.

\subsection{Canonical one-dimensional dynamics}

The canonical 1D model is typically derived by writing the Boussinesq equations in a rotated coordinate system aligned with a slope that is inclined at an angle $\theta$ above the horizontal \citep[e.g.,][]{garrett_boundary_1993}.
We here deviate from this approach by remaining in the un-rotated coordinates, which is a slightly more natural choice if the horizontal components of the turbulent momentum and buoyancy fluxes are neglected but yields equivalent dynamics.
Assuming no variations of the flow, pressure perturbation, or buoyancy perturbation in planes parallel to the slope (see Appendix~A for a more detailed derivation), we obtain
\begin{align}
    \pder{u}{t} - f \varv &= b\tan\theta + \pder{}{z}\left(\nu \pder{u}{z} \right)\label{eq:canonical-x},\\
    \pder{\varv}{t} + f u &= \pder{}{z}\left(\nu \pder{\varv}{z}\right)\label{eq:canonical-y},\\
    \pder{b}{t} + u N^2 \tan\theta &= \pder{}{z}\left[\kappa\left(N^2 + \pder{b}{z}\right)\right]\label{eq:canonical-b},
\end{align}
where $u$ is the cross-slope velocity, $\varv$~is the along-slope velocity, and $f$ is the (constant) inertial frequency.
As explained in Appendix~A,~$u$ is the horizontal projection of the cross-slope velocity as it would be defined in a fully rotated coordinate system, but we will still refer to it as the cross-slope velocity for simplicity.
We have split the total buoyancy~$B$ into a constant background stratification and a perturbation so that $B = N^2 z + b$.
Turbulent momentum and buoyancy transfer are represented by a diffusive closure with turbulent viscosity $\nu$ and turbulent diffusivity $\kappa$, related by the turbulent Prandtl number~$\mu = \nu / \kappa$.
We explore the consequences of using Rayleigh drag, a lower-order closure, in Appendix~C \citep[cf.,][]{callies_dynamics_2018,drake_abyssal_2020}.
The fluid satisfies no-slip and insulating boundary conditions at the bottom: $u = 0$, $\varv = 0$, and $\partial_z B = N^2 + \partial_z b = 0$ at $z = x \tan \theta = 0$, assuming (without loss of generality) that we apply these equations at $x = 0$.
At the upper boundary, we impose no stress and a fixed buoyancy flux $-\kappa N^2$: $\partial_z u = 0$, $\partial_z \varv = 0$, and $\partial_z b = 0$ at $z = H + x \tan \theta = H$ at $x = 0$.
The evolution is independent of $H$ if $H$ is large, in which case the domain can be considered semi-infinite.
Importantly, the assumption that the pressure perturbation does not vary in the cross-slope direction leaves only the projection of the buoyancy force in \eqref{eq:canonical-x}.

Numerical, analytical, and approximate solutions to these equations for both constant and bottom-enhanced $\kappa$ can be found in the literature \citep[e.g.,][]{garrett_boundary_1993,callies_restratification_2018}.
The system has a steady state, in which the turbulent buoyancy flux convergence or divergence is balanced by cross-slope advection.
This steady state is approached during both spin up and spin down first by rapid adjustment in the boundary layer, followed by a slow set up of a non-zero along-slope flow in the interior \citep{thorpe_current_1987,maccready_buoyant_1991,garrett_boundary_1993}.
Outside the boundary layer, the dominant balance in \eqref{eq:canonical-x} is $-f \varv = b \tan \theta$, so the along-slope flow and buoyancy perturbations evolve in lockstep. 
Combined with the other two equations, this yields
\begin{equation}\label{eq:diffusive-growth}
    (1 + S) \pder{b}{t} = \pder{}{z} \left(\kappa \left[ N^2 + (1 + \mu S) \pder{b}{z} \right] \right),
\end{equation}
implying that the adjustment of the far field is diffusive and thus slow \citep{maccready_buoyant_1991}.

Throughout the evolution, the vertically integrated buoyancy budget is
\begin{equation}\label{eq:canonical-b-vertical-int}
    \int_0^\infty \pder{b}{t} \; \text{d}z + UN^2\tan\theta = \kappa_\infty N^2,
\end{equation}
where $\kappa_\infty$ is the far-field diffusivity. 
This implies that the steady state is achieved by balancing the turbulent buoyancy flux into the water column with a net upslope transport $U = \kappa_\infty \cot\theta$.
During the transient, however, there is no explicit constraint on the cross-slope transport, and cross-slope transport in the boundary layer does not need to be returned above.
This canonical 1D model thus lacks a closed secondary circulation that could produce a more rapid adjustment of the along-slope flow than through slow diffusion.

\subsection{Canonical one-dimensional model in the planetary geostrophic framework}

By considering both the 1D and 2D dynamics in the PG limit, we can directly compare their inversion statements and clarify the role of transport through an explicit term in the equations.
The PG approximation assumes large horizontal scales and small Rossby numbers, rendering the tendency terms in the momentum equations negligible.
This approximation is reasonable for mixing-generated spin up in the abyss, but the tendency terms are crucial in Ekman arrest and spin down.
The simplified PG dynamics clearly illustrate the importance of constrained transport, however, which is ultimately key in both cases.

With the PG approximation applied, the canonical 1D equations \eqref{eq:canonical-x} to~\eqref{eq:canonical-b} become:
\begin{align}
    -f \varv &= b\tan\theta + \pder{}{z}\left(\nu \pder{u}{z} \right)\label{eq:canonical-x-pg},\\
    f u &= \pder{}{z}\left(\nu \pder{\varv}{z}\right)\label{eq:canonical-y-pg},\\
    \pder{b}{t} + u N^2\tan\theta &= \pder{}{z}\left[\kappa\left(N^2 + \pder{b}{z}\right)\right]\label{eq:canonical-b-pg}.
\end{align}
Given a buoyancy perturbation~$b$, the momentum equations \eqref{eq:canonical-x-pg} and \eqref{eq:canonical-y-pg} allow us to invert for the flow $(u, \varv)$, and the buoyancy is evolved in time through~\eqref{eq:canonical-b-pg}.
We define a streamfunction $\chi(z)$ such that $u = \partial_z \chi$, allowing us to cast the inversion as a single streamfunction equation.
Integrating~\eqref{eq:canonical-y-pg} from some level to $z = H$ yields
\begin{equation}\label{eq:v-from-chi}
    \pder{\varv}{z} = \frac{f}{\nu}(\chi - U).
\end{equation}
Differentiating~\eqref{eq:canonical-x-pg} and substituting $\partial_{z} \varv$ from~\eqref{eq:v-from-chi} yields the streamfunction inversion equation: 
\begin{equation}\label{eq:1dpg-inversion}
    \pder{^2}{z^2}\left(\nu\pder{^2\chi}{z^2}\right) + \frac{f^2}{\nu}(\chi - U) = -\pder{b}{z}\tan\theta.
\end{equation} 
The boundary conditions are that $\chi = 0$ and $\partial_z \chi = 0$ at $z = 0$ and $\chi = U$ and $\partial^2_{z} \chi = 0$ at $z = H$. 
Although not needed for the evolution, the along-slope flow can also be inferred from $\chi$ by integrating~\eqref{eq:v-from-chi} from the bottom up, using $\varv = 0$ at $z = 0$.

In these equations, the vertically integrated transport $U$ must be treated as an unknown ($U = \kappa_\infty\cot\theta$ applies in steady state only).
We must supplement~\eqref{eq:1dpg-inversion} with an additional boundary condition. 
Enforcing $\varv = 0$ at $z = 0$ in~\eqref{eq:canonical-x-pg} yields
\begin{equation}\label{eq:1dpg-extra-bc}
    \pder{}{z}\left(\nu\pder{^2\chi}{z^2}\right) = -b\tan\theta \quad \text{at} \quad z = 0,
\end{equation}
which closes the system and allows us to determine $U$ implicitly.
As we will see in the next section, however, this vertically integrated transport is constrained by the non-local context in 2D and 3D geometries and cannot evolve as freely as in these canonical 1D equations.

\subsection{Two-dimensional planetary geostrophic dynamics}

Consider the mixing-generated spin up of PG flow over the idealized 2D ridge sketched in Fig.~\ref{fig:sketchRidge}.
If the 1D model is to serve its purpose, then we should expect it to provide an accurate description of the local flow on the flanks of the ridge.
Continuity and symmetry imply that the vertically integrated cross-ridge transport within this domain must be zero, however, in contrast with the canonical model.
This simple example of a non-local constraint on transport illustrates a key piece of physics missing from the canonical 1D theory.

To make this comparison explicit, we consider the 2D PG equations for a fluid with depth $H(x)$,
\begin{align}
    -f\varv &= -\pder{p}{x} + \pder{}{z}\left(\nu \pder{u}{z}\right)\label{eq:2dpg-x}\\
    fu &= \pder{}{z}\left(\nu \pder{\varv}{z}\right)\label{eq:2dpg-y}\\
    \pder{p}{z} &= b\label{eq:2dpg-z}\\
    \pder{u}{x} + \pder{w}{z} &= 0\label{eq:2dpg-continuity}\\
    \pder{b}{t} + u \pder{b}{x} + w \left(N^2 + \pder{b}{z}\right) &= \pder{}{z}\left[\kappa\left(N^2 + \pder{b}{z}\right)\right]\label{eq:2dpg-buoyancy}
\end{align}
where $p$ is the pressure divided by a reference density and $w$ is the vertical velocity.
The boundary conditions are again an insulating and no-slip bottom, $N^2 + \partial_z b = 0$ and $u = \varv = 0$ at $z = -H$; a constant-flux and free-slip top $\partial_z b = 0$ and $\partial_z u = \partial_z \varv = 0$ at $z = 0$; and no normal flow across both boundaries, which together with $u = 0$ at $z = -H$ reduces to $w = 0$ at $z = -H$ and $z = 0$.

As before, we turn the momentum equations \eqref{eq:2dpg-x} to \eqref{eq:2dpg-continuity} into one streamfunction inversion.
Defining $\chi(x, z)$ such that $u = \partial_z \chi$ and $w = -\partial_x \chi$, we have
\begin{equation}\label{eq:2dpg-inversion}
    \pder{^2}{z^2}\left(\nu \pder{^2\chi}{z^2}\right) + \frac{f^2}{\nu}(\chi - U) = \pder{b}{x},
\end{equation}
where $U = \int_{-H}^0 u \; \text{d}z$ is the vertically integrated transport, a constant in $x$ by continuity.
The boundary conditions are similar to the 1D case: $\chi = 0$ and $\partial_z \chi = 0$ at $z = -H$ and $\chi = U$ and $\partial^2_z \chi = 0$ at $z = 0$.

The inversion equations \eqref{eq:1dpg-inversion} and \eqref{eq:2dpg-inversion} have the same form in 1D and 2D.
Under the assumption that $b$ does not vary in planes parallel to the slope, $\partial_x b = -\partial_z b \tan \theta$.
Continuity and symmetry over our 2D ridge (Fig.~\ref{fig:sketchRidge}), however, set the transport term to zero---whereas the canonical 1D model generally produces a time-varying $U \neq 0$.
This explicit difference between the two inversions causes qualitative differences between the 1D and 2D solutions, as seen in \citet{ruan_mixing-driven_2020} and further discussed below (Fig.~\ref{fig:spinupProfilesMu1}).
In general, the 1D dynamics are coupled to the barotropic vorticity equation via the vertically integrated transport terms.
The sinusoidal ridge considered here is a simple incarnation of this coupling in which the transport is always zero.
Although this choice of geometry is specific, it is not contrived; it should be possible to explain the dynamics over the ridge flanks with 1D theory.
The same principles still hold for asymmetric 2D geometries, where $U$ must be determined as part of the inversion but again is the result of a non-local constraint (see Appendix~B).

\subsection{Transport-constrained one-dimensional dynamics}

The analysis of the PG inversions in the previous section suggests that a 1D model must include an additional constraint on $U$ to faithfully reproduce local 2D dynamics.
The canonical 1D model \eqref{eq:canonical-x}~to~\eqref{eq:canonical-b} must therefore be modified to include another degree of freedom, with a natural choice being a vertically constant, time-varying pressure gradient $\partial_x P$.
This pressure gradient can accelerate a barotropic cross-slope flow~$u$ as needed to satisfy the transport constraint.
The transport-constrained 1D dynamics are then
\begin{align}
    \pder{u}{t} - f \varv &= -\pder{P}{x} + b\tan\theta + \pder{}{z}\left(\nu \pder{u}{z} \right)\label{eq:tc-x},\\
    \pder{\varv}{t} + f u &= \pder{}{z}\left(\nu \pder{\varv}{z}\right)\label{eq:tc-y},\\
    \pder{b}{t} + u N^2\tan\theta &= \pder{}{z}\left[\kappa\left(N^2 + \pder{b}{z}\right)\right]\label{eq:tc-b}, \\
    \int_0^H u \; \text{d}z &= U\label{eq:tc-U},
\end{align}
with $U$ prescribed. 
The tendency terms $\partial_t u$ and $\partial_t \varv$ are dropped if the PG approximation is applied.

As we will see in the solutions presented below, the transport constraint and barotropic pressure gradient are what allow the system to rapidly adjust in the interior.
Physically, the requirement that $U = 0$ forces any boundary layer transport to be returned in the interior (Fig.~\ref{fig:sketchSlope}).
This secondary circulation is not the same as the dipole in the diabatic circulation generated by bottom-intensified mixing; it is present even in the case of constant~$\kappa$ and acts on the entire column.
The interior cross-slope flow~$u$ is then turned by the Coriolis acceleration via~\eqref{eq:tc-y}, leading to rapid adjustment in the far-field along-slope flow~$\varv$.
This sets up geostrophic balance with the barotropic pressure gradient in the interior: $-f\varv = -\partial_x P$.
Classic Ekman spin up and spin down dynamics are now captured.

It should be noted that this geostrophic adjustment occurs instantaneously if the PG approximation is applied. 
The secondary circulation that sets up the barotropic along-slope geostrophic flow is therefore only implicit in the PG model and not part of the explicit streamfunction~$\chi$.

\begin{figure*}
    \centering
    \includegraphics[width=33pc]{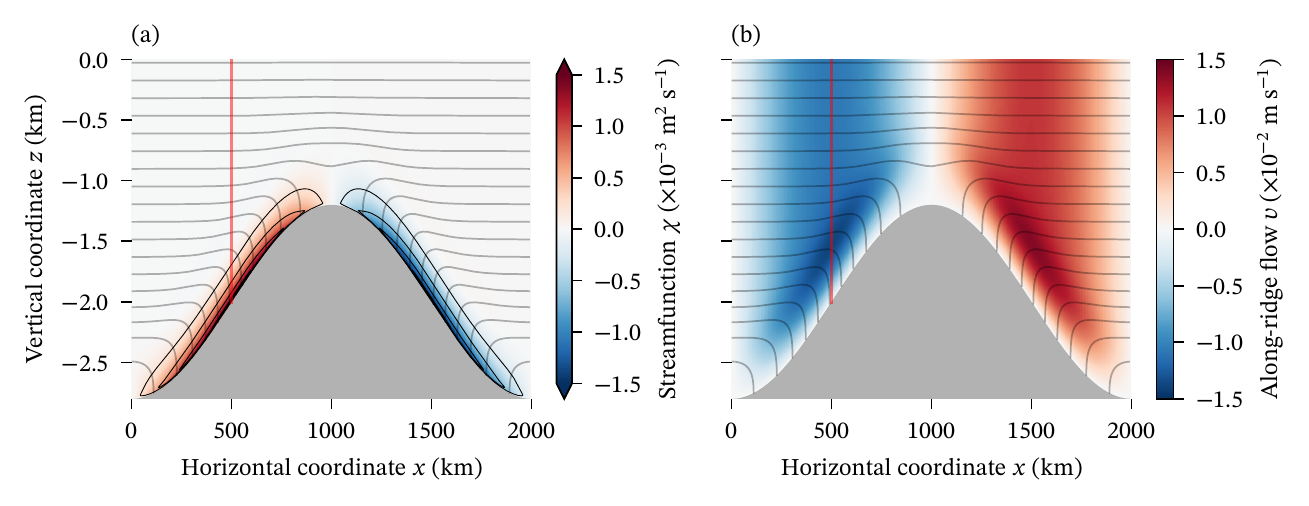}
    \caption{Flow fields in a 2D $\nu$PGCM simulation of mixing-generated spin up over the sinusoidal ridge sketched in Fig.~\ref{fig:sketchRidge}.
    Shown are (a)~the streamfunction~$\chi$ (shading and black contours) with positive values indicating counter-clockwise and negative values indicating clockwise flow and (b)~the along-ridge flow~$\varv$ (shading).
    The solution is shown after three years of spin up with bottom-intensified~$\kappa$ and~$\mu = 1$.
    The gray curves show isopycnals, and the red vertical lines show where 1D profiles are examined in Figs.~\ref{fig:spinupProfilesMu1} and~\ref{fig:spinupProfilesMu200}.
    }
    \label{fig:spinupRidge}
\end{figure*}

\section{Mixing-Generated Spin Up Over an Idealized Ridge}\label{s:spinup}

The modification to the 1D system described in the previous section enables it to capture the rapid spin up of interior flow encountered in 2D dynamics.
The canonical 1D model fails to do so.
To demonstrate this, we employ 1D and 2D numerical models to perform a mixing-generated spin-up experiment over the idealized symmetric ridge depicted in Fig.~\ref{fig:sketchRidge}.
For simplicity, we use the PG approximation for all models in this section, although subtle differences in spin up between PG and full models are noted in Appendix~D.

\subsection{Numerical models}

\begin{table}[b]
    \centering
    \begin{tabular}{l c c}
        \hline\hline
        Inertial frequency & $f$ & \SI{-5.5e-5}{\per\second}\\
        Far-field buoyancy frequency & $N$ & \SI{e-3}{\per\second}\\
        Far-field diffusivity & $\kappa_0$ & \SI{6e-5}{\meter\squared\per\second}\\
        Bottom-enhancement of diffusivity & $\kappa_1$ & \SI{2e-3}{\meter\squared\per\second}\\
        Decay scale of diffusivity & $h$ & 200 m\\
        Prandtl number & $\mu$ & 1 or 200\\
        \hline
    \end{tabular}
    \caption{Parameters used in the spin-up calculations, taken from \citet{ruan_mixing-driven_2020} and roughly corresponding to the Mid-Atlantic Ridge flank in the Brazil Basin.
    }
    \label{tab:params}
\end{table}

We solve the 2D PG system given by the inversion equation \eqref{eq:2dpg-inversion} and evolution equation \eqref{eq:2dpg-buoyancy} using terrain-following coordinates and second-order finite differences \citep[cf.,][]{callies_dynamics_2018}.
Model parameters and geometry are taken from \citet{ruan_mixing-driven_2020} to roughly match those of the Brazil Basin (Table~\ref{tab:params}), except that we enlarge the ridge to a more realistic size because the computational constraints from \citet{ruan_mixing-driven_2020} do not apply here.
Specifically, we take the domain height to be a sinusoid:
\begin{equation}
    H(x) = H_0 + A \cos \frac{2 \pi x}{L}
\end{equation}
with $H_0 = \SI{2}{\kilo\meter}$, $A = \SI{800}{\meter}$, and $L = \SI{2000}{\kilo\meter}$.
Mixing is represented by a bottom-intensified profile of turbulent diffusivity, 
\begin{equation}
    \kappa = \kappa_0 + \kappa_1 e^{-(z + H)/h},
\end{equation}
with parameters obtained from a fit to Brazil Basin observations \citep[][Table~\ref{tab:params}]{callies_restratification_2018}.
To reduce the impact of the upper boundary on the solution, we increase $H(x)$ uniformly by \SI{1}{\kilo\meter} compared to \citet{ruan_mixing-driven_2020} and apply $\partial_z b = 0$ rather than $N^2 + \partial_z b = 0$ at $z = 0$.
This ensures that isopycnals remain very nearly flat at the top of the domain, such that the PG evolution does not depend on the height of the domain.
Horizontal grid spacing is uniform at about 7~km, whereas vertical grid spacing follows Chebyshev nodes with resolution on the order of 0.1~m at $z=-H$ to comfortably resolve the boundary layers.
We time step the full buoyancy $B$ (rather than $b$) using a mixed implicit--explicit scheme and a time step of 10~days.
We refer to this model as the 2D~$\nu$PGCM.

We attempt to reproduce the 2D~$\nu$PGCM solution locally with the two 1D theories, using the local slope angle ($\theta \approx \num{2.5e-3}$ radians) and fluid depth ($H = 2$ km) at the center of the ridge flank ($x = \SI{500}{\kilo\meter}$, Fig.~\ref{fig:spinupRidge}).
The 1D models use the same numerical methods as the 2D~$\nu$PGCM to solve the inversion equation~\eqref{eq:1dpg-inversion} and evolution equation~\eqref{eq:canonical-b-pg} over a single column.
For the canonical case, the extra boundary condition given by~\eqref{eq:1dpg-extra-bc} is employed, whereas for the transport-constrained case $U = 0$ is specified.
The depth $H$ is large enough that upper-boundary effects do not affect the solution.
All models are initialized with $b = 0$, so that the total buoyancy is initially $B = N^2 z$.

\subsection{Results}

\begin{figure*}
    \centering
    \includegraphics[width=27pc]{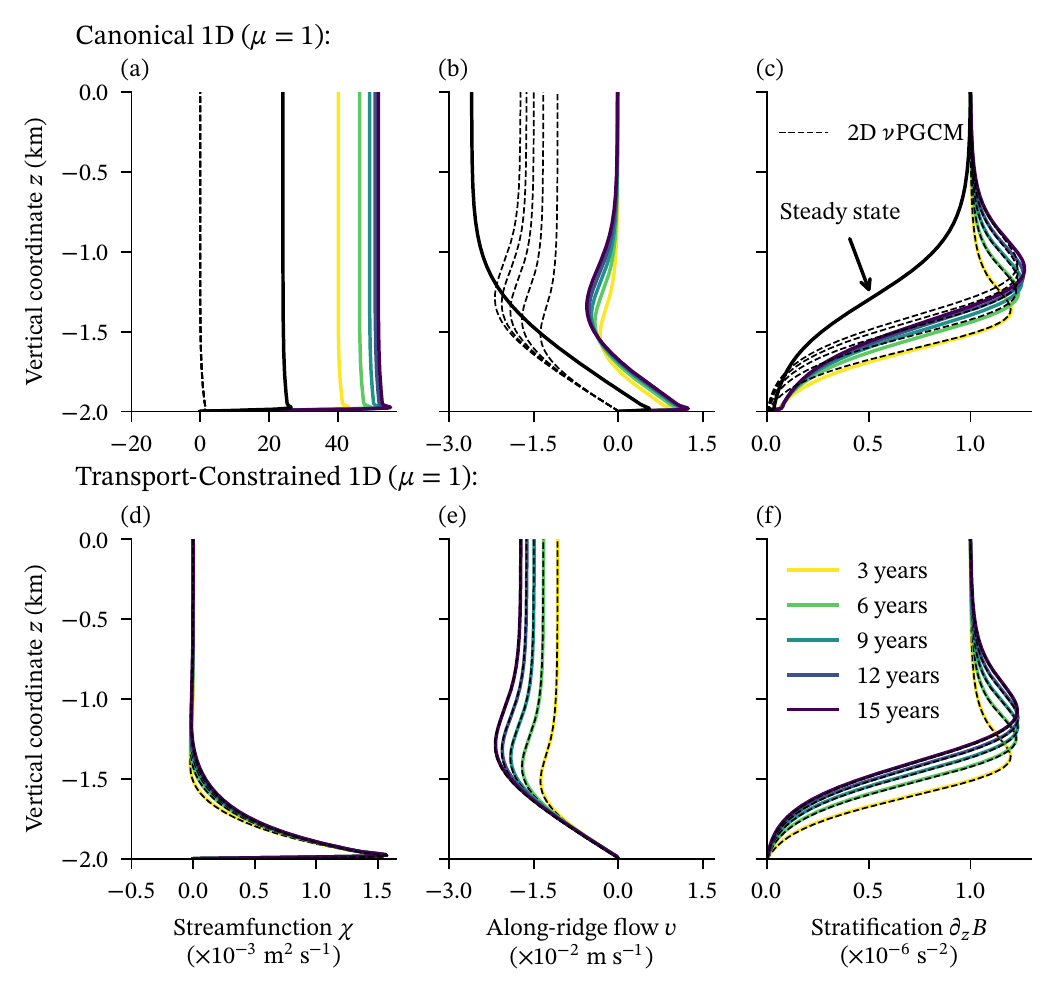}
    \caption{Comparison between the canonical and transport-constrained 1D solutions and their ability to capture the 2D simulation of mixing-generated spin up over a ridge.
    For all solutions, $\mu = 1$, and the profiles are taken at $x = \SI{500}{\kilo\meter}$ (red lines in Fig.~\ref{fig:spinupRidge}).
    Shown are the (a),~(d)~streamfunction~$\chi$, (b),~(e)~along-ridge flow~$\varv$, and (c),~(f)~stratification~$\partial_z B$.
    The first row (a--c) shows the canonical 1D solution (steady state in black) while the second row (d--f) shows the transport-constrained 1D solution.
    All panels include the 2D $\nu$PGCM solution (dotted) for comparison.
    }
    \label{fig:spinupProfilesMu1}
\end{figure*}

\begin{figure*}
    \centering
    \includegraphics[width=27pc]{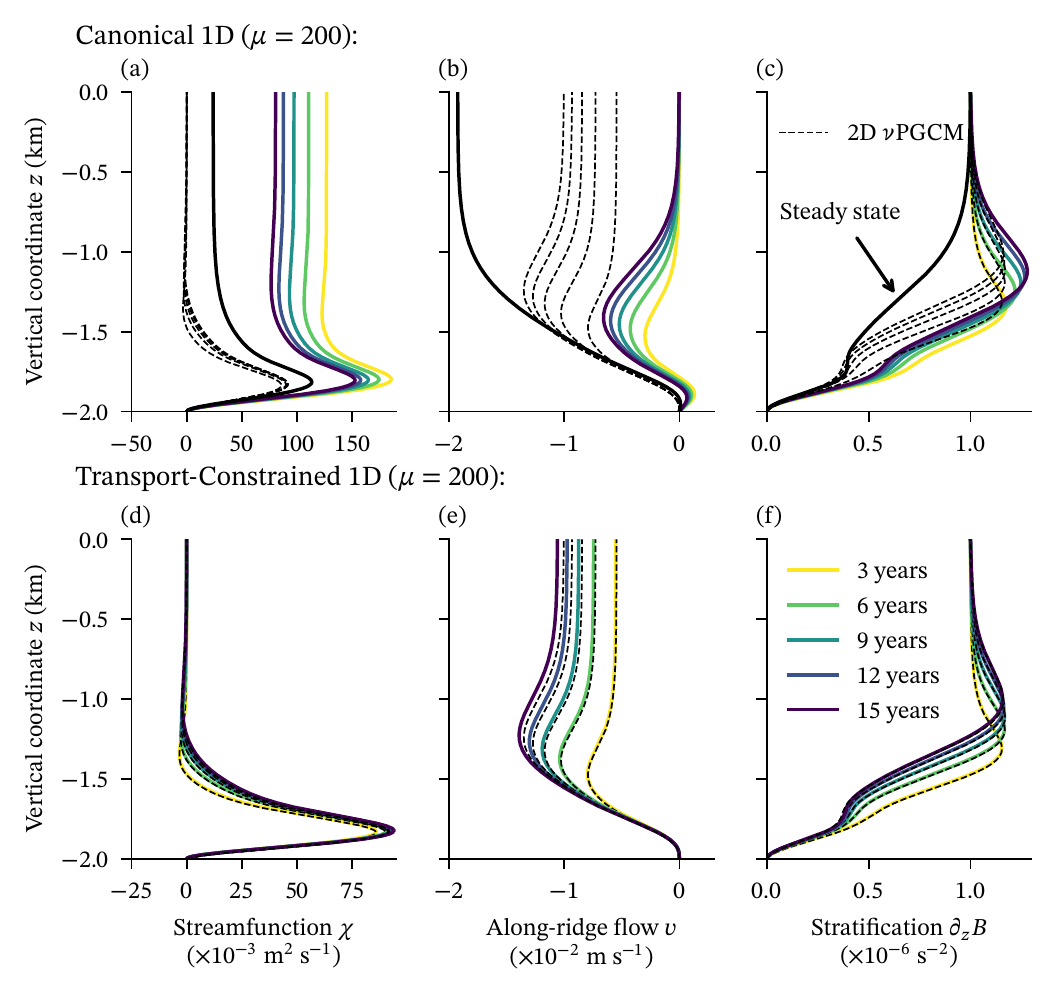} 
    \caption{Comparison between the canonical and transport-constrained 1D solutions and their ability to capture the 2D simulation of mixing-generated spin up over a ridge.
    For all solutions, $\mu = 200$, and profiles are taken at $x = \SI{500}{\kilo\meter}$ (red lines in Fig.~\ref{fig:spinupRidge}).
    Shown are the (a),~(d)~streamfunction~$\chi$, (b),~(e)~along-ridge flow~$\varv$, and (c),~(f)~stratification~$\partial_z B$.
    The first row (a--c) shows the canonical 1D solution (steady state in black) while the second row (d--f) shows the transport-constrained 1D solution.
    All panels include the 2D $\nu$PGCM solution (dotted) for comparison.}
    \label{fig:spinupProfilesMu200}
\end{figure*}

The insulating boundary condition at $z = -H$ leads to a buoyancy flux convergence and thus a positive buoyancy anomaly at the bottom, bending isopycnals into the ridge and spinning up a circulation (Fig.~\ref{fig:spinupRidge}).
Bottom-intensified mixing also produces buoyancy flux divergence above, causing isopycnals to bend up before plunging towards the slope.
Strong upwelling develops in a thin bottom boundary layer, broader and weaker downwelling occurs above, and a geostrophic along-slope flow emerges throughout the water column (Fig.~\ref{fig:spinupRidge}).
Our PG solutions are nearly identical to those of \citet{ruan_mixing-driven_2020}, who simulated the full primitive equations using the MITgcm (Appendix~D).

The canonical 1D theory fails to capture the evolution on the ridge flanks \citep[Fig.~\ref{fig:spinupProfilesMu1}a--c,][]{ruan_mixing-driven_2020}.
The canonical 1D theory predicts upslope flow in the bottom boundary layer that is an order of magnitude stronger than in the 2D system.
The 2D streamfunction differs substantially from the canonical 1D theory, which produces substantial net cross-slope transport.
The canonical 1D model predicts a diffusive progression of the along-slope flow into the interior, and substantial bottom stress induces the strong upslope Ekman transport.
By contrast, the 2D simulation's transport constraint leads to zero bottom stress in the along-slope flow because \eqref{eq:v-from-chi} implies $\partial_z \varv = 0$ at $z = -H$ when $U = 0$.
The buoyancy evolution is similar between the two models, except that the strong cross-slope flow in the canonical 1D solution maintains a stronger stratification in the bottom boundary layer.

In contrast with the canonical model, the transport-constrained 1D~model matches the results from the 2D~$\nu$PGCM very well (Fig.~\ref{fig:spinupProfilesMu1}d--f).
By enforcing the $U = 0$ constraint, we enable the streamfunction to match the 2D solution.
Additionally, the secondary circulation sets up a barotropic pressure gradient that allows the far-field along-slope flow to rapidly adjust rather than grow diffusively as in the canonical theory.
Finally, the two models yield nearly identical buoyancy profiles.
In both models, advection is negligible, and the buoyancy evolution is dominated by diffusion.
This is confirmed by separate simulations without the buoyancy advection terms, which yield very nearly identical solutions to those in Fig.~\ref{fig:spinupProfilesMu1} (not shown).

The transport-constrained 1D evolution equation only includes the cross-slope advection of the background buoyancy gradient $N^2 \tan \theta$, neglecting nonlinear transport terms.
A system in which advection plays a more dominant role in the evolution of buoyancy would be a more challenging test of the transport-constrained 1D model.
To achieve such a scenario, we increase the Prandtl number to $\mu = 200$ as a crude parameterization of baroclinic eddies \citep[e.g.,][]{rhines_homogenization_1982,greatbatch_parameterizing_1990,callies_restratification_2018,holmes_tracer_2019}.
The transport-constrained 1D~model still accurately describes the 2D~dynamics under these conditions (Fig.~\ref{fig:spinupProfilesMu200}d--f).
The increased Prandtl number thickens the boundary layer and strengthens the upwelling, which in turn maintains some of the stratification near the bottom boundary.
The far-field along-slope flow still adjusts rapidly, although the transport-constrained 1D model slightly over-predicts this evolution, caused by minor differences in the buoyancy field arising from the 2D advection missing in the 1D model.
The canonical 1D theory continues to fail miserably (Fig.~\ref{fig:spinupProfilesMu200}a--c).

\section{Spin Down and Ekman Arrest}\label{s:spindown}

As argued in the introduction, transport-constrained 1D~dynamics can also elucidate the interplay between Ekman arrest and spin down on a slope.
We ask how an initially barotropic along-slope flow~$V$, which is in geostrophic balance with a cross-slope pressure gradient, $fV = \partial_x P$, adjusts to the presence of a sloping boundary.
The current generates a cross-slope Ekman transport.
This Ekman transport has two effects: it acts on the cross-slope buoyancy gradient to produce buoyancy anomalies that slow down this transport, and, through the transport constraint, it produces a secondary circulation in the interior that spins down the initial along-slope flow~$V$.
Depending on the relative timescales of arrest and spin down, either the secondary circulation spins down $V$, or the Ekman transport is arrested before $V$ has been spun down, in which case the bottom becomes slippery and the flow persists.

\begin{figure*}
    \centering
    \includegraphics[width=27pc]{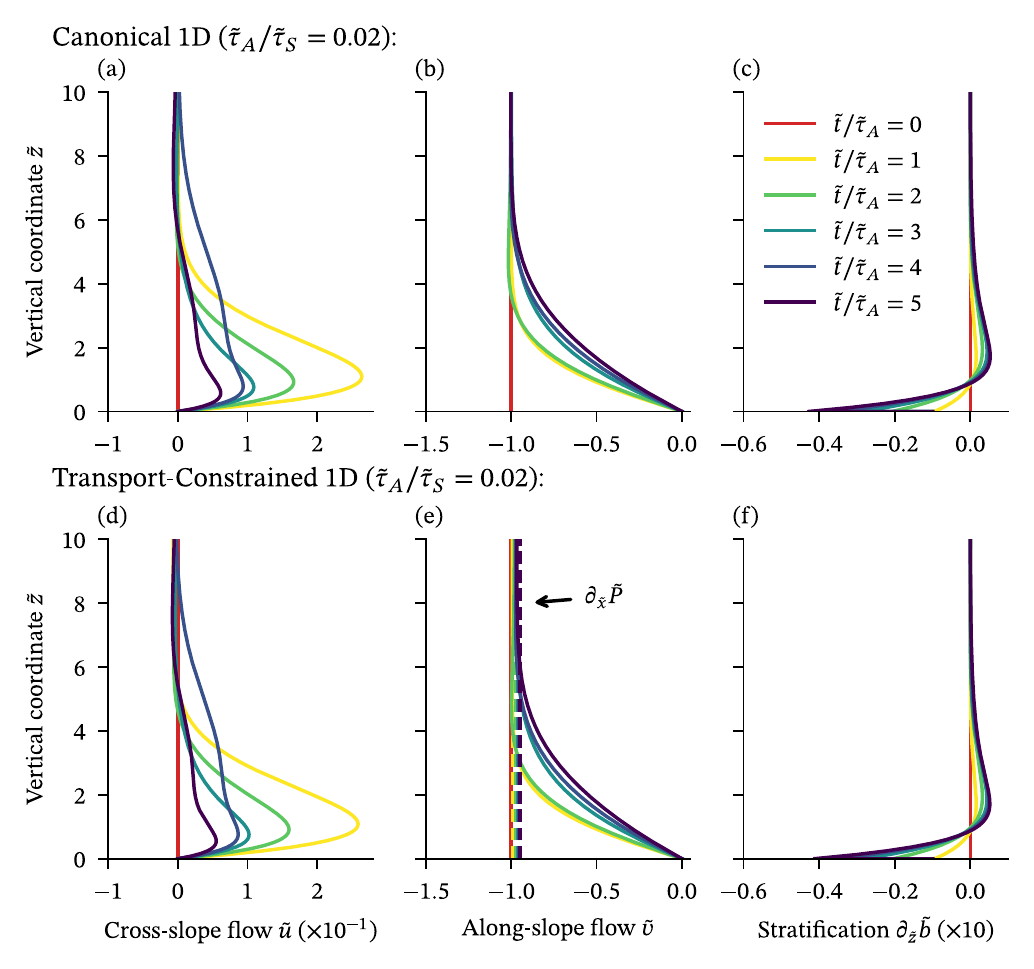}
    \caption{Comparison between the canonical and transport-constrained 1D simulations of spin down in a regime where Ekman arrest dominates.
    The Ekman number is $\Ek = \num{e-4}$, and the slope Burger number is $S = 0.5$, such that $\tilde \tau_A = 2$, $\tilde \tau_S = \num{e2}$, and $\tilde \tau_A / \tilde \tau_S = \num{2e-2}$.
    Shown are the (a),~(d)~cross-slope flow~$\tilde u$, (b),~(e)~along-ridge flow~$\tilde \varv$, and (c),~(f)~perturbation stratification~$\partial_{\tilde z} \tilde b$ in increments of Ekman arrest times.
    The first row (a--c) shows the canonical 1D solution, while the second row (d--f) shows the transport-constrained 1D solution.
    The barotropic pressure gradient $\partial_{\tilde x} \tilde P$ is shown in dashed lines in~(e) and held fixed at $-1$ in~(b).
    For clarity, only the first 10 Ekman layer depths are shown, but the full domain height is $H/\delta = \Ek^{-1/2} = 100$.
    }
    \label{fig:spindownRatioSmall}
\end{figure*}

This problem has been studied with the canonical 1D~model by imposing a $\partial_x P$ that balances the initial flow and is held fixed in time \citep{maccready_buoyant_1991,garrett_boundary_1993}.
Without a transport constraint, this model only contains the physics of Ekman arrest, with no mechanism for spinning down the interior flow other than slow diffusion.
The transport-constrained 1D model, in contrast, captures the secondary circulation and thus the physics of spin down.
In this model, $\partial_x P$ is allowed to change with time, as needed to satisfy the transport constraint $U = 0$.
In the following, we review the timescales for Ekman spin down and arrest and map out the parameter space using the transport-constrained 1D~model.

\subsection{Nondimensional one-dimensional equations}

To distill the dynamics down to its fundamental parameters, we nondimensionalize the 1D equations by setting
\begin{equation}
    t = T\tilde{t}, \quad z = \delta \tilde{z}, \quad u = V\tilde{u}, \quad \varv = V\tilde{\varv}, \quad b = \mathcal{B}\tilde{b}.
\end{equation}
We assume a constant viscosity~$\nu$ and set $\kappa = 0$ to focus on arrest and spin down without the effects of buoyancy diffusion \citep[cf.,][]{maccready_buoyant_1991}.
We choose an inertial timescale, the Ekman layer height scale, and a buoyancy scale corresponding to the buoyancy anomaly produced by cross-slope Ekman advection persisting for one inertial timescale:
\begin{equation}\label{eq:scalings}
    T = \frac{1}{f}, \quad \delta = \sqrt{\frac{\nu}{f}}, \quad \mathcal{B} = \frac{VN^2\tan\theta}{f}.
\end{equation}
With these scales, equations \eqref{eq:tc-x}~to~\eqref{eq:tc-U} become
\begin{align}
    \pder{\tilde u}{\tilde t} - \tilde{\varv} &= -\pder{\tilde P}{\tilde x} + S\tilde{b} + \pder{^2 \tilde u}{\tilde z^2} \label{eq:1dnd-x},\\
    \pder{\tilde\varv}{\tilde t} + \tilde{u} &= \pder{^2 \tilde \varv}{\tilde z^2} \label{eq:1dnd-y},\\
    \pder{\tilde b}{\tilde t} + \tilde{u} &= 0\label{eq:1dnd-b},\\
    \int_0^{H/\delta} \tilde{u} \; \text{d}\tilde{z} &= \tilde U = 0 \label{eq:1dnd-U},
\end{align}
where we set $\tilde U = 0$.
The nondimensional height of the domain may be written as $H/\delta = \Ek^{-1/2}$.
The model is thus fully characterized by two nondimensional parameters: the slope Burger number~$S = N^2 \tan^2 \theta / f^2$ and the Ekman number~$\Ek = \nu / f H^2$.
It is worth noting that with these choices in the nondimensionalization, the total stratification becomes
\begin{equation}\label{eq:nondim-strat}
    \pder{\tilde B}{\tilde z} = \frac{f\delta\cot\theta}{V} + \pder{\tilde b}{\tilde z},
\end{equation}
introducing a third nondimensional number, $f\delta\cot\theta/V$, a measure of the background stratification.
As a consequence of setting $\kappa = 0$, the flow's evolution is independent of this parameter.

\subsection{Spin down and Ekman arrest timescales}

Ekman spin down occurs when the geostrophic far-field along-slope flow $\tilde \varv = \pm 1$ is eroded by a secondary circulation.
First, a cross-slope Ekman transport of order unity ($\tilde u \sim \mp 1$ over $0 < \tilde z \lesssim 1$) is generated.
If the along-slope current had lateral structure, variations in this Ekman transport would produce convergences and divergences that would drive a secondary circulation.
Despite not capturing such lateral variations in the current, the transport-constrained 1D~model does produce this secondary circulation.
The convergence and divergence of the Ekman transport is delegated to $\tilde x \to \pm \infty$ and the constraint $\tilde U = 0$ ensures that all cross-slope Ekman transport is returned in the interior.
Being distributed uniformly over the domain of height~$H$, this cross-slope return flow has a magnitude $\tilde u \sim \pm \delta / H = \pm \Ek^{1/2}$.
With negligible friction in the interior, \eqref{eq:1dnd-y} implies that this return flow is turned into the along-slope direction by the Coriolis acceleration, $\partial_{\tilde t} \tilde \varv \approx - \tilde u \sim \mp \Ek^{1/2}$, spinning down the initial flow.
This implies a spin-down timescale
\begin{equation}
    \tilde{\tau}_S = \frac{1}{\sqrt{\Ek}}.
\end{equation}
In dimensional terms, this is $\tau_S = f^{-1}\Ek^{-1/2}$ and a classical result \citep[e.g.,][]{pedlosky_geophysical_1979}.
It is worth noting that quasi-geostrophic dynamics suggest that, in a system with a characteristic lateral length scale $L$, the vertical height scale~$H$ in this scaling would be the minimum of $fL/N$ (the ``Prandtl scale'') and the fluid depth \citep[e.g.,][]{holton_influence_1965,maccready_buoyant_1991}.

Ekman arrest, in contrast, involves the interaction of buoyancy forces with Ekman transport across a slope.
As before, the initial along-slope flow $\tilde \varv = \partial_{\tilde x} \tilde P \sim \pm 1$ induces a cross-slope Ekman transport of order unity.
The transport acts on the cross-slope buoyancy gradient through~\eqref{eq:1dnd-b}, generating a buoyancy anomaly of magnitude $\tilde b \sim \pm \tilde \tau$ over a timescale $\tilde \tau$.
The buoyancy force opposes the transport, ultimately neutralizing it once $S\tilde{b} \sim \partial_{\tilde x} \tilde P$ in \eqref{eq:1dnd-x} and  the near-bottom along-slope flow has been eliminated without requiring any change in $\partial_{\tilde x} \tilde P$.
This yields an arrest timescale of
\begin{equation}
    \tilde{\tau}_A = \frac{1}{S},
\end{equation}
or, in dimensional form, $\tau_A = (Sf)^{-1}$ \citep[e.g.,][]{rhines_boundary_1989}. 
As pointed out by \citet{maccready_buoyant_1991}, this scaling needs modification if $S \gtrsim 1$, a regime in which the Ekman transport cannot be assumed to persist at its original magnitude for the full time $\tilde \tau$. 
We only straddle this parameter regime and ignore the correction proposed by \citet{maccready_buoyant_1991} for simplicity.
This makes our analysis of the above scaling relevant for abyssal ridges ($S \sim \num{e-3}$) and some continental slopes and seamounts ($S \sim \num{e-1}$).

\begin{figure*}
    \centering
    \includegraphics[width=27pc]{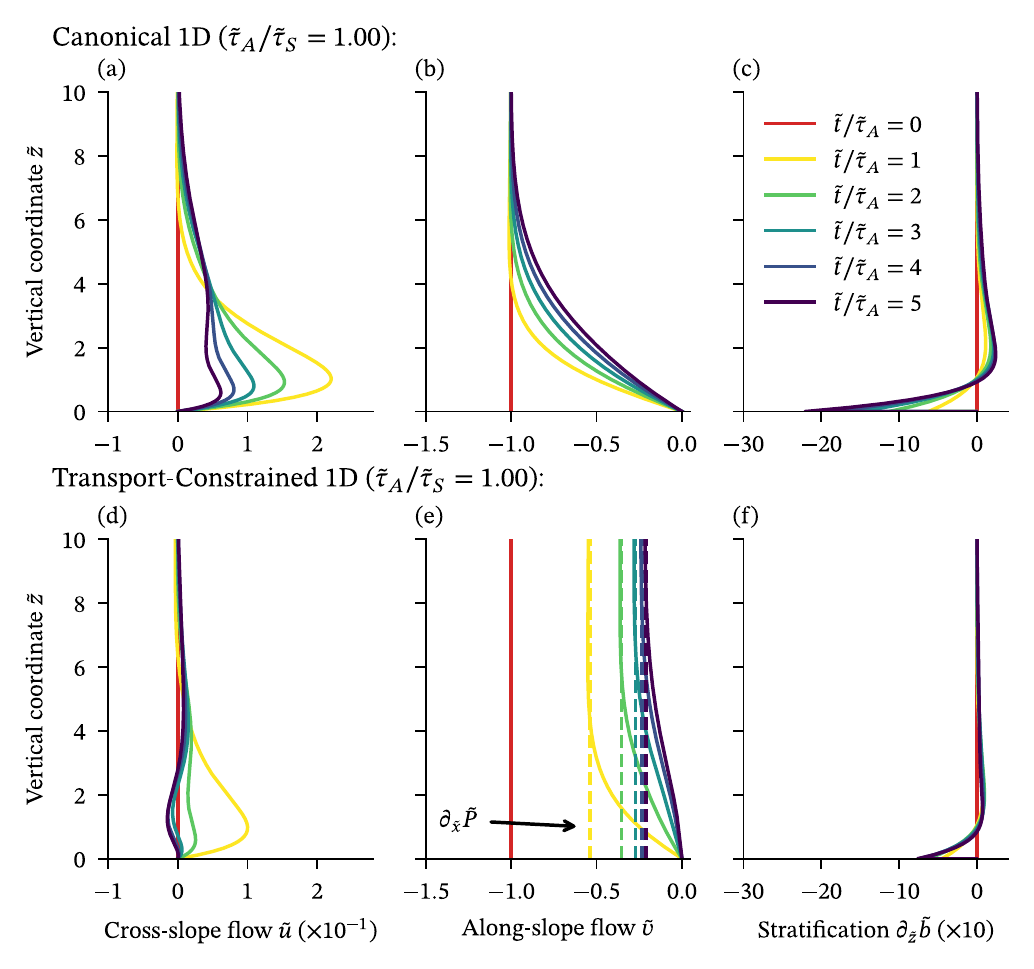}
    \caption{Comparison between the canonical and transport-constrained 1D simulations of spin down in a regime where spin down dominates.
    The Ekman number is $\Ek = \num{e-4}$, and the slope Burger number is $S = \num{e-2}$, such that $\tilde \tau_A = \num{e2}$, $\tilde \tau_S = \num{e2}$, and $\tilde \tau_A / \tilde \tau_S = 1$.
    Shown are the (a),~(d)~cross-slope flow~$\tilde u$, (b),~(e)~along-ridge flow~$\tilde \varv$, and (c),~(f)~perturbation stratification~$\partial_{\tilde z} \tilde b$ in increments of Ekman arrest times.
    The first row (a--c) shows the canonical 1D solution, while the second row (d--f) shows the transport-constrained 1D solution.
    The barotropic pressure gradient $\partial_{\tilde x} \tilde P$ is shown in dashed lines in~(e) and held fixed at $-1$ in~(b).
    For clarity, only the first 10 Ekman layer depths are shown, but the full domain height is $H/\delta = \Ek^{-1/2} = 100$.
    }
    \label{fig:spindownRatioBig}
\end{figure*}

Ekman spin down and arrest thus operate on different time scales.
If the spin-down timescale is short compared to the arrest timescale, the along-slope current is spun down before the Ekman transport is arrested.
Conversely, if the arrest timescale is short compared to the spin-down timescale, the Ekman transport is diminished before the current is spun down, and the arrested Ekman layer acts as an essentially slippery boundary condition for the persisting current.
This competition between the two processes is characterized by the ratio of their timescales \citep{garrett_boundary_1993}:
\begin{equation}\label{eq:spindown-ratio}
    \frac{\tilde{\tau}_A}{\tilde{\tau}_S} = \frac{\sqrt{\Ek}}{S}.
\end{equation}
When this ratio is large, we expect spin down; when it is small, we expect arrest.
These physics were identified by \citet{maccready_buoyant_1991}, but the canonical 1D~model employed there did not capture spin down and thus could not elucidate this competition explicitly.
As discussed in the introduction, the bulk model introduced in \citet{chapman_deceleration_2002} was capable of representing both processes but lacked a direct connection to the full equations of motion.

\subsection{Numerical results}

\begin{figure}
    \centering
    \includegraphics[width=19pc]{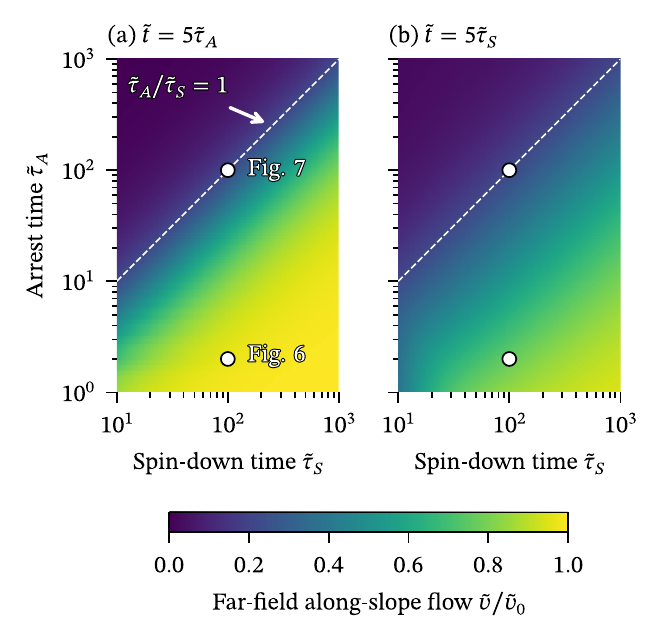}
    \caption{Competition between spin down and Ekman arrest in the transport-constrained 1D model.
    Colors show the fraction of the initial far-field along-slope flow~$\tilde \varv$ remaining after (a) five arrest times and (b) five spin-down times for a wide range of spin-down timescales~$\tilde \tau_S = \Ek^{-1/2}$ and arrest timescales~$\tilde \tau_A = 1/S$.
    }
    \label{fig:spindownGrid}
\end{figure}

We now explore the competition between spin down and Ekman arrest across the parameter space $(S, \Ek)$, solving equations \eqref{eq:1dnd-x}~to~\eqref{eq:1dnd-U} numerically using second-order finite differences over a grid of $2^9$~to~$2^{11}$ Chebyshev nodes (depending on $\Ek$) and a Crank--Nicolson timestepping scheme with a timestep of $\Delta \tilde t = \min\{\tilde\tau_A/100, \tilde\tau_S/100\}$.
As mentioned above, the depth of the domain depends on the Ekman number through $H/\delta = \Ek^{-1/2}$, so more nodes are required for smaller~$\Ek$.
We initialize all simulations with $\tilde{b} = 0$, $\tilde{u} = 0$, and a barotropic geostrophic flow $\tilde{\varv} = -1$, so as to induce upwelling in the Ekman layer.
This choice is made without loss of generality: downwelling solutions induced by $\tilde \varv = 1$ are equivalent due to symmetry of the system with $\kappa = 0$.
The along-slope flow must balance $\partial_{\tilde x} \tilde P$ so that, initially, we have $\partial_{\tilde x} \tilde P = \tilde \varv = -1$.
To compare with the canonical 1D theory, we hold $\partial_{\tilde x} \tilde P = -1$ fixed and drop the transport constraint \eqref{eq:1dnd-U} as in \citet{maccready_buoyant_1991}.
In the transport-constrained model, on the other hand, $\partial_{\tilde x} \tilde P$ is allowed to change in time, such that the extra constraint \eqref{eq:1dnd-U} can be satisfied.

We begin with a case in which Ekman arrest occurs before the interior flow is spun down.
With $S = 0.5$ and $\Ek = \num{e-4}$ ($H/\delta = 100$), the arrest and spin-down timescales are $\tilde \tau_A = 2$ and $\tilde \tau_S = \num{e2}$.
Their ratio is $\tilde{\tau}_A / \tilde{\tau}_S = 0.02$, so Ekman arrest is about 50~times faster than spin down.
This parameter regime might occur on the slopes of a typical seamount or on the continental slope.
Both the canonical and transport-constrained 1D~models capture Ekman arrest, so they should produce similar results in this regime.
Indeed, after five arrest times, the two model solutions show the same qualitative behavior (Fig.~\ref{fig:spindownRatioSmall}).
In both models, the Ekman transport decays with time.
The along-slope flow adjusts in the Ekman layer and shows a hint of diffusion into the interior in both cases, although the interior geostrophic flow is also spun down by a few percent in the transport-constrained 1D~model.
The stratification is enhanced by upwelling, although in the very bottom Ekman layer the models yield large negative perturbations to the stratification due to our choice of $\kappa=0$.
Depending on one's choice of nondimensional background stratification in \eqref{eq:nondim-strat}, this could lead to gravitationally unstable solutions.
This unphysical result was also encountered by \citet{maccready_buoyant_1991}, and subsequent studies used more sophisticated turbulence parameterizations to analyze the problem in the presence of convection \citep[e.g.,][]{trowbridge_asymmetric_1991,maccready_slippery_1993,brink_buoyancy_2010}.

As we move into a parameter regime where spin down becomes important, the two models diverge (Fig.~\ref{fig:spindownRatioBig}).
With $S = \num{e-2}$ and $\Ek = \num{e-4}$ ($H/\delta = 100$), the arrest and spin-down timescales are $\tilde \tau_A = \tilde \tau_S = \num{e2}$.
Spin down is now as important as Ekman arrest.
About 80\% of the original geostrophic flow is eroded in the transport-constrained model after five Ekman arrest times, whereas the interior along-slope flow (by design) remains fixed at $-1$ in the canonical model.
The rapid spin down of the geostrophic flow in the transport-constrained model also leads to much weaker Ekman transport and therefore smaller stratification changes in the boundary layer.
These results are in contrast with \citeauthor{chapman_deceleration_2002}'s (\citeyear{chapman_deceleration_2002}) model, which suggested a more prominent role of Ekman arrest in this parameter regime.
This quantitative difference might stem from differences in turbulence closures; \citet{chapman_deceleration_2002} used linear bottom drag, allowing his model to reach a non-trivial steady state.
A more direct comparison between the two models can be achieved by employing the same turbulence closures in the transport-constrained 1D model, but that is beyond the scope of this paper.
As the slope Burger number is further reduced to $S \sim \num{e-3}$ (and $\Ek$ held fixed), a value typical for abyssal ridge flanks such as in Fig.~\ref{fig:sketchRidge}, spin down becomes strongly dominant over Ekman arrest.

To assess how accurately the ratio $\tilde \tau_A / \tilde \tau_S$ captures the competition between Ekman spin down and arrest in the transport-constrained model, we compute solutions with $\tilde \tau_A$ and $\tilde \tau_S$ varied over multiple orders of magnitude.
The simple ratio of Ekman arrest time to spin down time captures the dynamics of the far-field along-slope flow remarkably well (Fig.~\ref{fig:spindownGrid}).
After five arrest times, $\tilde t = 5 \tilde \tau_A$, simulations with a larger $\tilde{\tau}_A/\tilde{\tau}_S$ have smaller geostrophic flows than those with smaller ratios (Fig.~\ref{fig:spindownGrid}a).
If $\tilde{\tau}_A/\tilde{\tau}_S > 1$, the interior flow has been almost completely spun down at $\tilde t = 5 \tilde \tau_A$, whereas for $\tilde{\tau}_A/\tilde{\tau}_S < 0.1$, the interior flow is almost entirely preserved at $\tilde t = 5 \tilde \tau_A$ because Ekman arrest has prevented spin down.
Spin down is not entirely prevented, however.
After five spin-down times, $\tilde t = 5 \tilde \tau_S$, the geostrophic current is substantially eroded, even when $\tilde{\tau}_A/\tilde{\tau}_S < 0.1$ (Fig.~\ref{fig:spindownGrid}b).

\section{Discussion}\label{s:discussion}

The transport-constrained model does not generally allow for a steady state.
Part of the attraction of the canonical 1D model has been that it achieves a steady-state balance between buoyancy advection and diffusion.
It has become apparent here, however, that this comes at the expense of implying a peculiar choice for the cross-slope mass transport: $U$ is implicitly chosen such that the barotropic cross-slope pressure gradient is eliminated.
This choice is clearly incorrect in our example of a simple 2D ridge.
Our discussion therefore challenges the significance of the steady transport $U = \kappa_\infty \cot \theta$ of the canonical model.
If the canonical steady state was for some reason desired, one could recover it by setting $U = \kappa_\infty \cot \theta$ in the transport-constrained 1D model, but it is not clear to us how that might be justified.\footnote{Even if $\kappa_\infty = 0$, such that the canonical model has a steady state with $U = 0$, the evolution of the two models remains dramatically different because $U \neq 0$ in the canonical model before the steady state is reached.}
Instead, we argue that the transport~$U$ is the result of coupling with the non-local part of the dynamics and that achieving a steady state must also involve these non-local dynamics.
The transport-constrained model thus encourages a reconsideration of the interaction between the boundary layer and interior dynamics.
Boundary layer theory can be used to clarify the physics of this interaction, a topic we are planning to discuss in a separate manuscript.

The lack of a steady state also complicates discussions of the effectiveness of boundary mixing \citep{garrett_role_1990,garrett_boundary_1993,garrett_isopycnal_2001}. While advective restratification tends to be weaker with the transport constraint \citep[cf.,][]{ruan_mixing-driven_2020}, a full discussion of this point must involve non-local effects that balance the net lightening in the 1D column, such that a steady state can be reached.

The dependence of the transport-constrained 1D~model on the domain height~$H$ is worth clarifying.
The spin-down physics discussed in Section~\ref{s:spindown} depend explicitly on~$H$ because the magnitude of the cross-slope return flow that develops in response to the Ekman transport depends on how deep a water column this return flow is distributed over.
The spin-down timescale $\tau_S$ is thus proportional to~$H$ (or the Prandtl scale if the current has lateral structure on a scale similar to or smaller than the deformation radius).
In contrast, the PG dynamics discussed in Section~\ref{s:spinup} are independent of~$H$ as long as isopycnals remain flat at the top of the domain.
This is because the geostrophic adjustment of the along-slope current occurs instantaneously if the momentum tendencies are dropped.
The actual rate of this adjustment, which does depend on~$H$, becomes immaterial in the PG limit.

In general PG dynamics, the vertically integrated transport arises from the coupling between columnar baroclinic 1D inversions and the barotropic vorticity equation.
The same is true in 2D, but the barotropic dynamics reduce to either $U = 0$ or an explicit formula for $U$ (see Appendix~B).
Thinking of the dynamics in this way, in conjunction with boundary layer theory, is both conceptually and computationally advantageous. 
Extended to 3D in future work, this approach allows for new insight into the role of the bottom boundary layer in the dynamics of the abyssal circulation.
The theory presented in this paper, however, does not lend itself to making claims about the large-scale context, and we do not make any effort to do so.

Throughout this work, we have relied on simple representations of turbulent momentum and buoyancy fluxes, certainly not giving justice to the complexity of turbulence in bottom boundary and stratified mixing layers.
Even in idealized spin-down scenarios, turbulence can be generated by a mix of shear, gravitational, symmetric, and centrifugal instabilities \citep{wenegrat_centrifugal_2020}.
Furthermore, we have ignored the presence of small-scale topography, which excites the strong internal-wave field that produces bottom-intensified turbulence \citep[e.g.,][]{nikurashin_mechanism_2011}, as well as baroclinic eddies, which might help restratify abyssal mixing layers \citep{callies_restratification_2018}.
More sophisticated turbulence parameterizations can be added to the transport-constrained equations, or the transport constraint can be added to local three-dimensional calculations in slope-aligned coordinates that resolve the turbulence \citep[e.g.,][]{wenegrat_submesoscale_2018,callies_restratification_2018,ruan_evolution_2019,wenegrat_centrifugal_2020,ruan_evolution_2021}.
In spin-down calculations, for example, a more faithful description of the turbulent dynamics would reintroduce the asymmetry between downwelling- and upwelling-favorable currents.
Despite this added complexity, however, the transport constraint and its consequences for rapid adjustment should remain important in many circumstances.

Under what circumstances does the canonical model remain accurate?
One might hope that it does if $\Ek^{1/2} \gg S$, so that Ekman arrest quickly halts the spin down that is not captured by the canonical model.
No matter how rapidly Ekman arrest occurs, however, we find that the transport-constrained 1D model still spins down the interior flow eventually (Fig.~\ref{fig:spindownGrid}b).
Conservatively, the canonical theory should thus be restricted to times $f t \ll \Ek^{-1/2}$, although the lifetime can be extended if Ekman arrest is fast enough to slow down the spin down process.
In any case, this argument renders the canonical steady state meaningless and implies that the canonical 1D model is never valid under the PG approximation, in which spin down is instantaneous.

Equipped with the transport-constrained model, one should revisit previous results that were based on canonical 1D dynamics.
In addition to the spin-down problem, in which slow diffusion is replaced by a rapid adjustment through a secondary circulation, several other topics might warrant reconsideration, for example:
\begin{enumerate}

    \item Motivated by observations over the East Pacific Rise, \citet{thompson_abyssal_1996} integrated the canonical equations starting from rest and with bottom-intensified mixing, very similar to the calculations presented in Section~\ref{s:spinup}.
    They found bottom-intensified along-slope currents and inferred transports comparable to deep western boundary currents.
    Transport-constrained dynamics, however, produce flow that instead decays towards the bottom (compare Fig.~\ref{fig:spinupProfilesMu1}b,e).
    While it remains unclear what happens in the presence of a planetary vorticity gradient, when interior meridional flow must be attended by vortex stretching, it is apparent that the canonical solutions should be considered less than definitive.
    
    \item The mean flows discussed in \citet{callies_restratification_2018} would similarly be altered by a transport constraint. 
    Cross-slope transport in the bottom boundary layer is weaker when the integrated transport is constrained (e.g., Fig.~\ref{fig:spinupProfilesMu1}a,d), implying that restratification by mean flows is even weaker than implied by the canonical model employed in \citet{callies_restratification_2018}.
    The conclusion that baroclinic eddies are crucial in enhancing the stratification in abyssal mixing layers is thus robust, as confirmed in \citet{ruan_mixing-driven_2020}, where submesoscale eddies were found to dominate in a 3D model with constrained transport. 
    The utility of the steady solutions to the canonical equations presented in \citet{callies_restratification_2018}, however, is called into question.
    
    \item \citet{benthuysen_friction_2012} examined the effects of boundary mixing on the potential vorticity (PV) of the fluid during the spin down of an initial along-slope current.
    In the canonical 1D model that they employed, the interior current is diffusively eroded until a non-trivial steady flow is reached, as described by \citet{maccready_buoyant_1991}.
    \citet{benthuysen_friction_2012} found that the initial flow direction relative to the steady flow determines whether PV is injected or extracted from the fluid.
    If this study were revisited with the transport-constrained 1D model, the qualitative behaviour of the flow would be altered, at least if spin down dominates over Ekman arrest (as in Figs.~\ref{fig:spindownRatioBig}).
    The conclusion that PV fluxes primarily depend on the direction of the initial current, however, relies only on Ekman buoyancy flux physics and is likely robust.

\end{enumerate}

\section{Conclusions}\label{s:conclusions}

Recent work has highlighted the role that abyssal mixing layers play in the circulation of the abyssal ocean \citep{ferrari_turning_2016,de_lavergne_consumption_2016,mcdougall_abyssal_2017,holmes_ridges_2018,callies_dynamics_2018,drake_abyssal_2020}.
A starting point for understanding these dynamics of a stratified, rotating fluid overlying an inclined seafloor has been the canonical 1D theory first developed by \citet{phillips_flows_1970} and \citet{wunsch_oceanic_1970}.
We have shown here, however, that the choice to set the cross-slope pressure gradient to zero in these dynamics eliminates important physics.
If instead a constraint is imposed on the vertically integrated cross-slope transport, which can be thought of as arising from the non-local context of the 1D column, and a barotropic cross-slope pressure gradient is allowed, rapid spin up and spin down of the interior along-slope flow can be captured.
With this transport constraint, a secondary cross-slope circulation can develop in the 1D framework, even if there are no lateral variations in the flow, and act on the interior flow.
These modified 1D dynamics accurately capture the mixing-generated spin up over an idealized 2D ridge, where the canonical 1D dynamics fail.
It can be hoped that these transport-constrained 1D dynamics can serve as a more reliable cornerstone for building a theory of the abyssal circulation than the canonical 1D system.

Capturing the Ekman spin down of an interior current, the transport-constrained 1D model can also be used to study the competition between spin down and Ekman arrest in a unified framework.
We have presented the simplest model of this competition, employing a constant viscosity and no buoyancy diffusion, in which previous expectations are exactly matched.
For $S \ll 1$, the competition is described completely by the ratio of spin-down and arrest timescales $\tau_A/\tau_S = \Ek^{1/2}/S$ \citep{maccready_buoyant_1991,garrett_boundary_1993}.
A more detailed exploration of these dynamics, with more realistic turbulence closures plugged into the transport-constrained model or with a transport constraint imposed on turbulence-resolving simulations, is left to future work.

\datastatement
The numerical models for all the simulations presented here are hosted at \url{https://github.com/hgpeterson/nuPGCM}.

\acknowledgments
We thank Henri Drake, Xiaozhou Ruan, and Paola Cessi for useful discussions of this work.
We also thank the two anonymous reviewers for their insightful comments on the manuscript.

\clearpage
\appendix[A]

\appendixtitle{1D Model in Coordinates Aligned with the Slope and Gravity}

\begin{figure*}
    \centering
    \noindent\includegraphics[width=27pc]{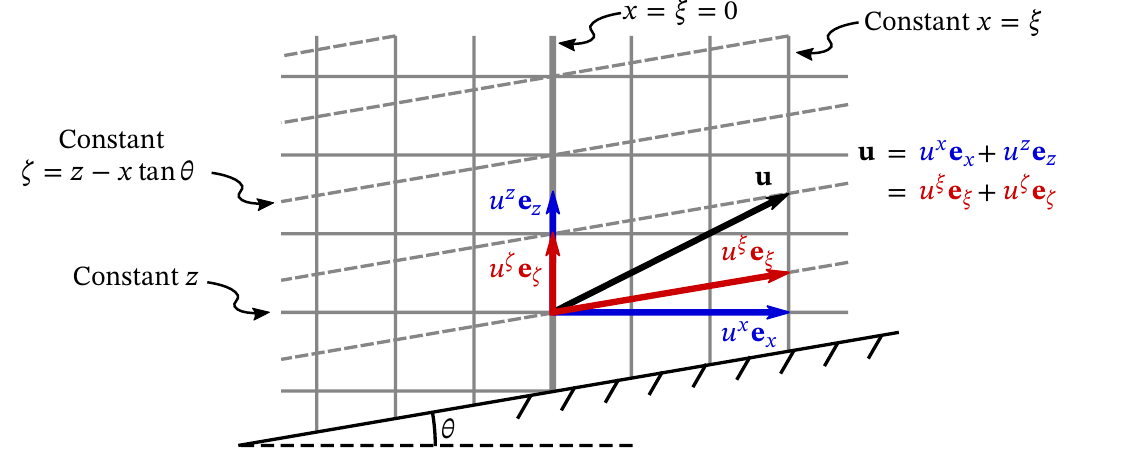}
    \caption{Sketch of the coordinates aligned with the slope and gravity as used in the 1D model.
    The covariant basis vector of coordinate~$j$ is denoted by~$\vec{e}_j$, and the corresponding contravariant component of the velocity vector is denoted by~$u^j$, such that $\vec{u} = u^j \vec{e}_j$ (summation implied).
    }
    \label{fig:1Dcoordinates}
\end{figure*}

\noindent Here we derive the 1D model by transforming into a coordinate system in which coordinate lines are aligned with the slope and with the direction of gravity (Fig.~\ref{fig:1Dcoordinates}). 
This coordinate system is a more natural choice than the often-used fully rotated coordinate system if the horizontal components of the turbulent momentum and buoyancy fluxes are neglected from the outset. If the turbulence is roughly isotropic, this neglect is consistent with the assumption of a small aspect ratio made to drop inertial terms in the vertical momentum equation.

The hydrostatic Boussinesq equations in Cartesian coordinates $(x, y, z)$, with $z$ aligned with gravity, read
\begin{align}
    \pder{u^x}{t} + \vec{u} \cdot \nabla u^x - f u^y &= -\pder{p}{x} + \pder{}{z} \left( \nu \pder{u^x}{z} \right), \\
    \pder{u^y}{t} + \vec{u} \cdot \nabla u^y + f u^x &= -\pder{p}{y} + \pder{}{z} \left( \nu \pder{u^y}{z} \right), \\
    b &= \pder{p}{z}, \\
    \pder{u^x}{x} + \pder{u^y}{y} + \pder{u^z}{z} &= 0, \\
    \pder{b}{t} + \vec{u} \cdot \nabla b + u^z N^2 &= \pder{}{z} \left[ \kappa \left( N^2 + \pder{b}{z} \right) \right],
\end{align}
where the velocity components are written using superscripts rather than as $u$, $\varv$, and $w$ as in the main text.
The superscripts indicate contravariant components, and this tensor notation helps keep the notation clear as we transform into the non-Cartesian coordinates.
We now define a new coordinate system $(\xi, \eta, \zeta)$ such that $\zeta = 0$ at the sloping boundary (Fig.~\ref{fig:1Dcoordinates}):
\begin{equation}
    \xi = x, \quad \eta = y, \quad \zeta = z - x \tan \theta.
\end{equation}
This is analogous to terrain-following coordinates but for an infinite slope and no horizontal upper boundary \citep[cf.,][]{callies_dynamics_2018}.
The contravariant velocity components under this coordinate transformation are then
\begin{equation}
    u^\xi = u^x, \quad u^\eta = u^y, \quad u^\zeta = u^z - u^x \tan \theta,
\end{equation}
and the partial derivatives transform as
\begin{equation}
    \pder{}{x} = \pder{}{\xi} - \tan \theta \pder{}{\zeta}, \quad \pder{}{y} = \pder{}{\eta}, \quad \pder{}{z} = \pder{}{\zeta}.
\end{equation}
Hydrostatic balance thus implies that
\begin{equation}
    -\pder{p}{x} = -\pder{p}{\xi} + \tan \theta \pder{p}{\zeta} = -\pder{p}{\xi} + b \tan \theta,
\end{equation}
so that the hydrostatic Boussinesq equations in this new coordinate system read
\begin{align}
    \pder{u^\xi}{t} + \vec{u} \cdot \nabla u^\xi - f u^\eta &= -\pder{p}{\xi} + b \tan \theta + \pder{}{\zeta} \left( \nu \pder{u^\xi}{\zeta} \right), \\
    \pder{u^\eta}{t} + \vec{u} \cdot \nabla u^\eta + f u^\xi &= -\pder{p}{\eta} + \pder{}{\zeta} \left( \nu \pder{u^\eta}{\zeta} \right), \\
    b &= \pder{p}{\zeta}, \\
    \pder{u^\xi}{\xi} + \pder{u^\eta}{\eta} + \pder{u^\zeta}{\zeta} &= 0, \\
    \pder{b}{t} + \vec{u} \cdot \nabla b + u^\xi N^2 \tan \theta + u^\zeta N^2 &= \pder{}{\zeta} \left[ \kappa \left( N^2 + \pder{b}{\zeta} \right) \right].
\end{align}
Neglecting all variations in $\xi$ and $\eta$, except for the barotropic pressure gradient $\partial_x P$ if desired, implies that $u^\zeta = 0$ by continuity, and the equations simplify to
\begin{align}
    \pder{u^\xi}{t} - f u^\eta &= -\pder{P}{x} + b \tan \theta + \pder{}{\zeta} \left( \nu \pder{u^\xi}{\zeta} \right), \\
    \pder{u^\eta}{t} + f u^\xi &= \pder{}{\zeta} \left( \nu \pder{u^\eta}{\zeta} \right), \\
    \pder{b}{t} + u^\xi N^2 \tan \theta &= \pder{}{\zeta} \left[ \kappa \left( N^2 + \pder{b}{\zeta} \right) \right].
\end{align}
Since $u^\xi = u^x$, $u^\eta = u^y$, and $\partial_\zeta = \partial_z$, these are equivalent to~\eqref{eq:canonical-x} to~\eqref{eq:canonical-b} (with $\partial_x P = 0$) and \eqref{eq:tc-x} to~\eqref{eq:tc-b} in the main text.
We note that $u^\xi = u^x$ is the horizontal projection of the cross-slope velocity as it would be defined in a fully rotated coordinate system.
This is because the basis vector~$\vec{e}_\xi = \vec{e}_x + \tan \theta \, \vec{e}_z$ does not have unit length (Fig.~\ref{fig:1Dcoordinates}).

\clearpage
\appendix[B]

\appendixtitle{Calculation of the Cross-Ridge Transport for General Topography}

\begin{figure}
    \centering
    \noindent\includegraphics[width=19pc]{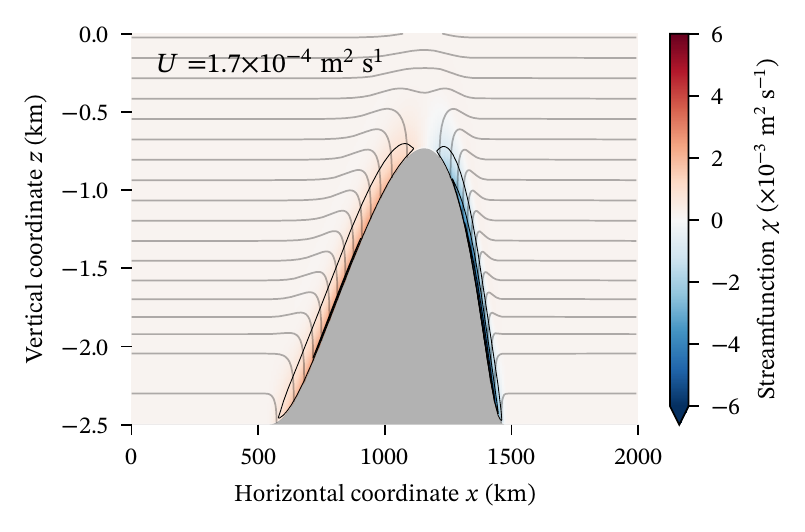}
    \caption{Mixing-generated spin up over an asymmetric ridge, showing net transport~$U \approx \SI{1.7e-4}{\meter\squared\per\second}$.
    The streamfunction (shading and black contours) is shown at 3 years, with positive values indicating counter-clockwise flow and negative values clockwise flow.
    The gray curves show isopycnals.
    }
    \label{fig:spinupRidgeAsym}
\end{figure}

\noindent For symmetric 2D bottom topography such as in Fig.~\ref{fig:sketchRidge}, it is immediately clear by continuity and symmetry that the vertically integrated cross-ridge flow~$U$ must vanish.
Similarly, if the depth~$H$ vanishes anywhere in the 2D domain, $U = 0$ everywhere follows by continuity.
For general topography in a 2D periodic domain, however, we need to compute~$U$ along with the PG inversion~\eqref{eq:2dpg-inversion}.
We here show how this can be done and illustrate the procedure with a solution for mixing-generated spin up over an asymmetric 2D ridge (Fig.~\ref{fig:spinupRidgeAsym}).

First, it is useful to split the streamfunction~$\chi$ into two components:
\begin{equation}\label{eq:chi-split}
    \chi = \chi^b + U\chi^U.
\end{equation}
The buoyancy component~$\chi^b$ is defined as solving
\begin{equation}
    \pder{^2}{z^2}\left(\nu \pder{^2 \chi^b}{z^2}\right) + \frac{f^2}{\nu} \chi^b = \pder{b}{x}
\end{equation}
with the boundary conditions $\chi^b = 0$ at both $z = -H$ and $z = 0$. 
The transport component~$\chi^U$ instead solves
\begin{equation}
    \pder{^2}{z^2}\left(\nu \pder{^2 \chi^U}{z^2}\right) + \frac{f^2}{\nu} \chi^U = \frac{f^2}{\nu},
\end{equation}
with the boundary conditions $\chi^U = 0$ at $z = -H$ and $\chi^U = 1$ at $z = 0$, such that \eqref{eq:chi-split} solves the inversion equation~\eqref{eq:2dpg-inversion} and satisfies the boundary conditions $\chi = 0$ at $z = -H$ and $\chi = U$ at $z = 0$. 
Note that both $\chi^b$ and $\chi^U$ are independent of~$U$ and can be calculated without its knowledge.

To obtain a formula for $U$, we follow a similar approach as in the classic ``Island Rule'' \citep[e.g.,][]{pedlosky_circulation_1997}.
We begin by taking the $x$-mean, denoted by $\langle\,\cdot\,\rangle$, of the $x$-momentum equation~\eqref{eq:2dpg-x} at $z = 0$, which gives
\begin{equation}
    -f\langle\varv\rangle - \left\langle\pder{}{z}\left(\nu \pder{u}{z}\right)\right\rangle = 0 \quad \text{at} \quad z = 0.
\end{equation}
Applying the definition of the streamfunction and the relation \eqref{eq:v-from-chi}, this can be written as
\begin{equation}
    \bigg\langle\overline{\frac{f^2}{\nu}(\chi - U)}\bigg\rangle + \left\langle\pder{}{z}\left(\nu \pder{^2 \chi}{z^2}\right)\right\rangle = 0 \quad \text{at} \quad z = 0,
\end{equation}
where $\overline{(\cdot)} = \int_{-H}^0 (\cdot) \; \text{d}z$.
Substituting~\eqref{eq:chi-split} and solving for $U$ yields
\begin{equation}\label{eq:U-computation}
    U = -\frac{\left\langle\pder{}{z}\left(\nu \pder{^2 \chi^b}{z^2}\right)\right\rangle_{z=0} + \Big\langle\overline{\frac{f^2}{\nu}\chi^b}\Big\rangle}{\left\langle\pder{}{z}\left(\nu \pder{^2 \chi^U}{z^2}\right)\right\rangle_{z=0} + \Big\langle\overline{\frac{f^2}{\nu}(\chi^U - 1)}\Big\rangle},
\end{equation}
thus completing the solution to~\eqref{eq:1dpg-inversion}.

To showcase the calculation using~\eqref{eq:U-computation}, we perform a simulation of mixing-generated spin up over the asymmetric ridge in Fig.~\ref{fig:spinupRidgeAsym}.
We obtain $U \approx \SI{1.7e-4}{\meter\squared\per\second}$ after three years of spin up.

\clearpage
\appendix[C]

\appendixtitle{Spin Up with Rayleigh Drag}

\noindent In studies of mixing-generated spin up in the abyss, the turbulent transport of momentum has been parameterized using Rayleigh drag by \citet{callies_dynamics_2018} and \citet{drake_abyssal_2020}.
Here we briefly show the consequences of such a closure in the context of the transport-constrained 1D theory.

The momentum and continuity equations for the 2D PG system with Rayleigh drag take the form 
\begin{align}
    -f\varv &= \pder{p}{x} - ru,\\
    fu &= -r\varv, \vphantom{\pder{p}{x}}\\
    \pder{p}{z} &= b,\\
    \pder{u}{x} + \pder{w}{z} &= 0,
\end{align}
where $r$ is a friction parameter.
The lower order of these equations compared with \eqref{eq:2dpg-x}~to~\eqref{eq:2dpg-continuity} reduces the number of boundary conditions that we may apply: we only require no-normal flow at the bottom and top boundaries.
As above, we define a streamfunction such that $\partial_z \chi = u$, yielding the inversion equation
\begin{equation}
    \frac{f^2 + r^2}{r}\pder{^2\chi}{z^2} = -\pder{b}{x},
\end{equation}
with boundary conditions $\chi = 0$ at $z = -H$ and $\chi = U$ at $z = 0$.
Notice that, in contrast to the case with Fickian momentum transfer, the streamfunction response to a buoyancy gradient is not localized in~$z$.
There is no height scale other than the domain height.
As in Appendix~B, the streamfunction can be split into buoyancy and transport components to obtain a formula for~$U$:
\begin{equation}
    U = -\left\langle\pder{\chi^b}{z}\right\rangle_{z = 0}\Bigg/\left\langle\pder{\chi^U}{z}\right\rangle_{z = 0}.
\end{equation}

Let us now compare this with the 1D system.
In a slope-aligned coordinate system and with a barotropic cross-slope pressure gradient included, the PG momentum equations with Rayleigh drag take the form
\begin{align}
    -f \varv &= -\pder{P}{x} + b\tan\theta - r u,\\
    f u &= -r \varv,
\end{align}
or, as a streamfunction equation,
\begin{equation}
    \frac{f^2 + r^2}{r} \pder{^2\chi}{z^2} = \pder{b}{z}\tan\theta,
\end{equation}
with boundary conditions $\chi = 0$ at $z = 0$ and $\chi = U$ at $z = H$.
Again, the inversion is equations are equivalent in 2D and the transport-constrained 1D system.
The lack of an additional height scale applies to the transport-constrained 1D model as well, which means that solutions depend strongly on the domain height~$H$.
This also means that the limit $H \to \infty$ is no attainable in the transport-constrained model with Rayleigh drag.
This is clear from the vertical integral of the momentum equations, which yields 
\begin{equation}
    \frac{f^2 + r^2}{r} U = -H \pder{P}{x} + \int_0^H b\tan\theta \; \mathrm{d}z.
\end{equation}
The limit $H \to \infty$ thus requires $\partial_x P \to 0$, but then the transport~$U$ cannot be specified separately.

As with Fickian diffusion, the transport-constrained 1D model better captures the 2D solution.
Rayleigh friction applies throughout the whole water column, however, causing return flow to spread across the full domain.
This leads to errors in both 1D models due to their slope-aligned coordinate system.

\clearpage
\appendix[D]

\appendixtitle{Comparison Between PG and Non-PG Transport-Constrained 1D Solutions}

\begin{figure*}
    \centering
    \noindent\includegraphics[width=27pc]{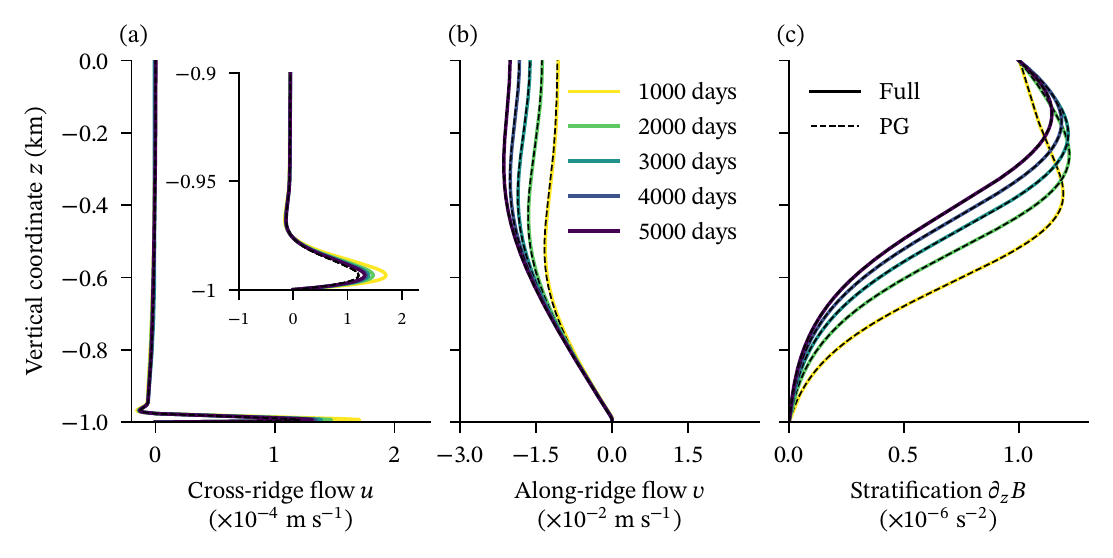}
    \caption{Comparison between PG and full transport-constrained 1D solutions for mixing-generated spin up.
    For all solutions, parameters are as in \citet{ruan_mixing-driven_2020} (i.e.~Table~\ref{tab:params} with $\mu = 1$, $\theta = \SI{2.5e-3}{}$, and $H = \SI{1}{\kilo\meter}$).
    Shown are the (a)~cross-slope flow~$u$, (b),~along-slope flow~$\varv$, and (c)~stratification~$\partial_{z} B$.
    Solid lines denote full solutions while dotted lines show PG solutions.
    The transport-constrained 1D PG model matches \citet{ruan_mixing-driven_2020} remarkably well (cf., their Fig.~4), with the full model capturing fast variations in Ekman transport [inset of panel~(a)].}
    \label{fig:spinupProfilesPGvsFull}
\end{figure*}

\noindent In the main text, we argue that the PG approximation is sufficient for describing the dynamics of mixing-generated spin up over an idealized ridge.
Additionally, we claim that our PG solutions match those of \citet{ruan_mixing-driven_2020}, who solved the 2D primitive equations.
For full transparency, we here show a comparison between the transport-constrained 1D dynamics with and without momentum tendency terms included (Fig.~\ref{fig:spinupProfilesPGvsFull}).

To directly compare with the solutions in \citet{ruan_mixing-driven_2020}, we use a domain height of $H = \SI{1}{\kilo\meter}$ and show the solutions in intervals of 1000~days.
The slope-aligned coordinate system in the 1D theory makes it difficult to reproduce their results with boundary conditions applied on a horizontal upper boundary.
To minimize boundary layer effects at the upper boundary, we therefore retain a constant buoyancy flux $-\kappa N^2$ as in the main text, which leads to slight differences in the upper \SI{200}{\meter} from \citet{ruan_mixing-driven_2020}.

Overall, the transport-constrained 1D PG model predicts mixing-generated spin up that is nearly identical to the 2D primitive equation solution shown in \citet{ruan_mixing-driven_2020} (cf., their Fig.~4).
The only substantial difference is that the Ekman transport in the PG system instantaneously adjusts and remains roughly constant throughout the 5000~day spin up, whereas the full system produces initially larger and subsequently decreasing cross-slope flow (Fig.~\ref{fig:spinupProfilesPGvsFull}a).
This arises because the initial buoyancy field does not satisfy the bottom boundary condition, so the initial adjustment is faster than the inertial timescale.

\clearpage
\bibliographystyle{ametsocV6}
\bibliography{biblio}

\end{document}